\documentclass[a4paper,11pt]{article}
\usepackage{pos}
\usepackage{cleveref}

\title{Conserving Lattice Gauge Theory for Finite Systems}

\author*[a]{Alexander Rothkopf}

\affiliation[a]{Faculty of Science and Technology, University of Stavanger, NO-4036 Stavanger, Norway}

\emailAdd{alexander.rothkopf@uis.no}

\abstract{In this contribution I discuss a recent proposal of a novel action for lattice gauge theory for finite systems, which accommodates non-periodic spatial boundary conditions. Drawing on the summation-by-parts formulation of finite differences and finite volume strategies of computational electrodynamics, an action is constructed that implements the proper integral form of Gauss' law and exhibits an inherently symmetric energy momentum tensor, all while realizing automatic ${\cal O}(a)$ improvement. Its central ingredients are illustrated using Abelian gauge theory as example.}

\FullConference{%
 The 38th International Symposium on Lattice Field Theory, LATTICE2021
  26th-30th July, 2021
  Zoom/Gather@Massachusetts Institute of Technology
}

\begin{document}
\maketitle

\section{Motivation}

The powerful combination of lattice gauge theory and Monte-Carlo methods in Euclidean time has over the past decade revealed many vital properties of strongly interacting matter under extreme conditions. One example is the phase structure of QCD at finite temperature and more recently also at small to moderate Baryo-chemical potential, which has been explored in detail (for recent works see e.g. \cite{Borsanyi:2021hbk,Borsanyi:2021sxv,Bazavov:2020bjn,HotQCD:2019xnw}). Such studies of the phase diagram are motivated in large part by heavy-ion collision experiments carried out at the LHC at CERN and formerly RHIC at BNL. Phenomenological models of locally thermalized nuclear matter in the form of relativistic hydrodynamics, amended by fist principles input, such as the QCD equation of state from lattice QCD, have been successful in interpreting the experimental findings in such collisions of heavy nuclei \cite{Busza:2018rrf,Jaiswal:2016hex}.

One still outstanding puzzle however emerged in the study of intermediate and small systems, in which tantalizing hints for the existence of a quark-gluon-plasma (QGP) were observed. These signals are similar to those in a large systems collision, even though the conditions for a transition of nuclear matter to the high temperature phase are much less favorable. One line of inquiry in this context asks, whether the small spatial size of these systems may mimic effects that in a large system can be ascribed to the existence of a strongly interacting QGP.

Recent exploratory work in the continuum and on the lattice has revealed that the presence of a finite volume and non-trivial boundary conditions can indeed lead to significant deviations from the infinite volume results in a region that may be of relevance for phenomenology. In Ref.\cite{Mogliacci:2018oea} the authors investigated the extreme example of free thermal scalar field theory, which due absence of any form of screening, probes the geometry on all length scales. It was shown that restricting the extent of one or more spatial directions lead to deviations from the Stefan-Boltzmann behavior in a fashion that reduces the pressure, a change similar to that induced by finite interactions. The more directions become constrained, the stronger the effect, so that in a box geometry of finite extent in all three spatial directions a reduction of the pressure by around 10\% was encountered at a value of $T L \approx 20$, and an even more significant reduction by 40\% at $T L \approx 3-5$. A $T L \approx 20$ is similar to the geometry encountered in a $A+A$ collision at LHC, while $T L \approx 3-5$ corresponds to what one might argue arises in a $p+A$ collisions. 

Obviously free scalar theory may not provide relevant insight for strongly interacting matter, which is why the results of Ref.\cite{Kitazawa:2019otp} are very interesting. They reveal that reducing the physical length of a single spatial axis (which led to the weakest deviation in the case of scalar theory) of a purely gluonic system to values between $1<L T <2$ leads to significant changes in the pressure associated with that direction, even flipping the sign of the pressure. While a first important step, this study still relied on periodic boundary conditions and only restricted one of the spatial directions. 

Both studies hint at the relevance of finite volume effects in nuclear matter under current experimental conditions. In order to systematically explore these effects of a genuine finite extent of the fireball, created in a relativistic heavy-ion collision, we thus need to formulate lattice gauge theory in a way that accommodates non-periodic (e.g. Dirichlet) boundary conditions. (Note that in a gauge theory such boundary conditions must of course be formulated in terms of gauge invariant expressions on the boundary of the spatial volume.)

A second motivation is recent work on the heavy quark potential in classical statistical gauge theory, where in Ref.\cite{Lehmann:2020fjt} for the first time the screened real-part of that potential was successfully computed. The take home message of that study was that charges in classical LGT are not introduced by the evaluation of the Wilson loop observable but have to the treated explicitly in a modified Gauss' law. 

In turn one may ask whether there are alternatives to evaluating the Wilson loop and the answer is in the affirmative. Ref.\cite{Yanagihara:2018qqg} showed how to extract the potential from the gauge invariant stress tensor, the spatial components of the energy momentum tensor. The discretization prescription formulated in the following section sets out to provide a lattice gauge theory which in its energy momentum tensor accurately reproduces the symmetric field lines of the simplest possible system, that of a static charge and anti-charge, already on the classical level and at the same time fulfills Gauss' law.

Taken together our goal here is to construct a lattice gauge theory, which accommodates systems with non-trivial boundary conditions and those where translational invariance is not given, such as in the presence of external sources.

\section{Two Challenges}

The first challenge I wish to focus on is the treatment of non-periodic boundary conditions. Let us inspect the standard Wilson action $S=\frac{1}{g^2}\sum_{x}\big[ \sum_i 2\frac{a_s}{a_t} {\rm ReTr} [P^{1\times 1}_{0i}-1] -\sum_{i,j}\frac{a_t}{a_s} {\rm ReTr} [P^{1\times 1}_{ij}-1]\big]$ expressed in terms of plaquettes
\begin{align}
\nonumber P^{1\times1}_{\mu\nu,x}&=U_{\mu,x} U_{\nu,x+a_\mu{\hat\mu}}U^\dagger_{\mu,x+a_\nu{\hat\nu}}U^\dagger_{\nu,x} = e^{i a_\mu a_\nu \tilde F_{\mu\nu,x}}+{\cal O}(a^2),\\
&\tilde F_{\mu\nu}=\Delta^{\rm F}_\mu A_{\nu,x} - \Delta^{\rm F}_\nu A_{\mu,x} + i [ A_{\mu,x},A_{\nu,x}],\label{eq:Waction}
\end{align} 
which themselves are constructed of link variables $U_{\mu,x}={\rm exp}[i a_\mu A_{\mu,x}]$. As is well known, the above corresponds to a discretization of the field strength tensor in terms of forward finite differences $\Delta^{\rm F}_\mu \phi(x)= (\phi(x+a_\mu \hat\mu) - \phi(x) )/a_\mu$, with $a_\mu$ denoting the lattice spacing and $\hat \mu$ the unit vector in the $\mu$ direction. In order to distill the essence of the challenge and not be encumbered by the technical difficulties of non-Abelian gauge theory let us explore this discretization in the Abelian case first.

\begin{figure}
\centering
\includegraphics[scale=0.23]{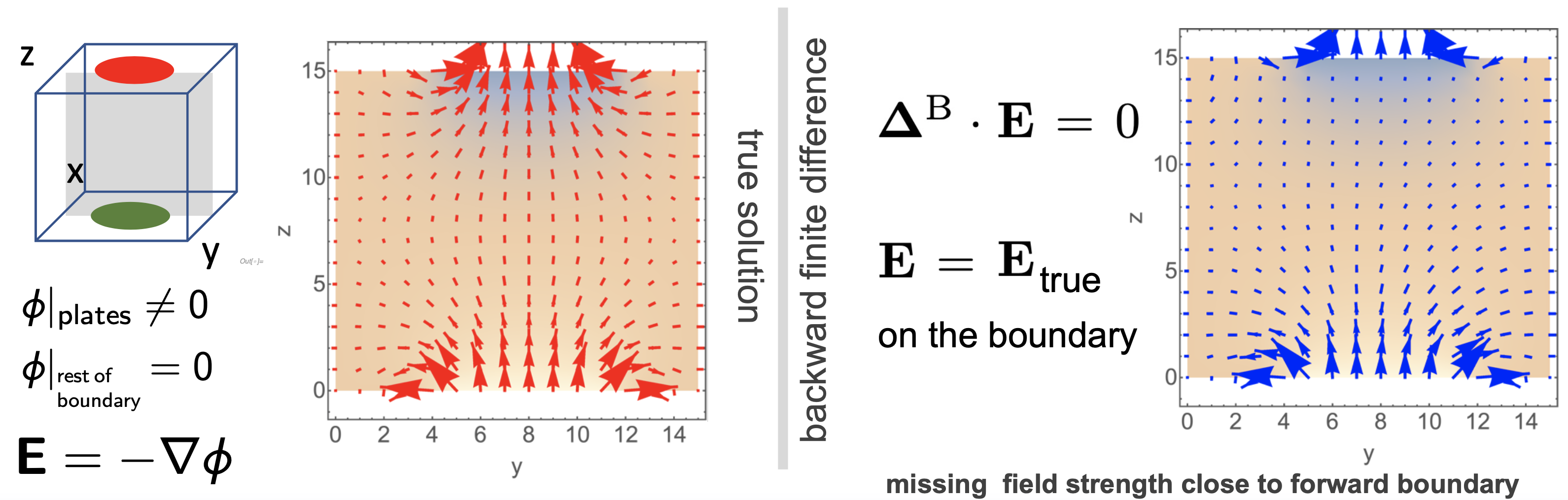}
\caption{(left) sketch of the geometry and the true solution (red) of the field lines for the example of a circular capacitor at finite potential. (right) backward finite difference solution (blue) of Gauss' law with the true solution as boundary condition.}\label{Fig:AbelianGaussCap1}
\end{figure}

The quantization of gauge theories requires the formulation of a Hamiltonian picture, which is amended by a constraint in the form of Gauss' law and which singles out physical states within the Hilbert space. Therefore let me focus on the solutions of the discretized Gauss' law in the presence of non-periodic boundary conditions. 

On the left of \cref{Fig:AbelianGaussCap1} I show as an example the geometry of a circular capacitor held at a finite potential. The classical potential in the interior may be computed from solving the Poisson equation, the corresponding field lines ensue from taking the negative gradient. If this solution, referred to as "true" in the following, is supplied on the boundary for the solution of Gauss' law with backward finite differences (corresponding to the forward FD of the Wilson action) one obtains the solution shown in blue on the right. It is easy to see that while the backward FD can produce field strength at the backward boundary it fails to do so at the forward boundary.

\begin{figure}
\centering
\includegraphics[scale=0.25]{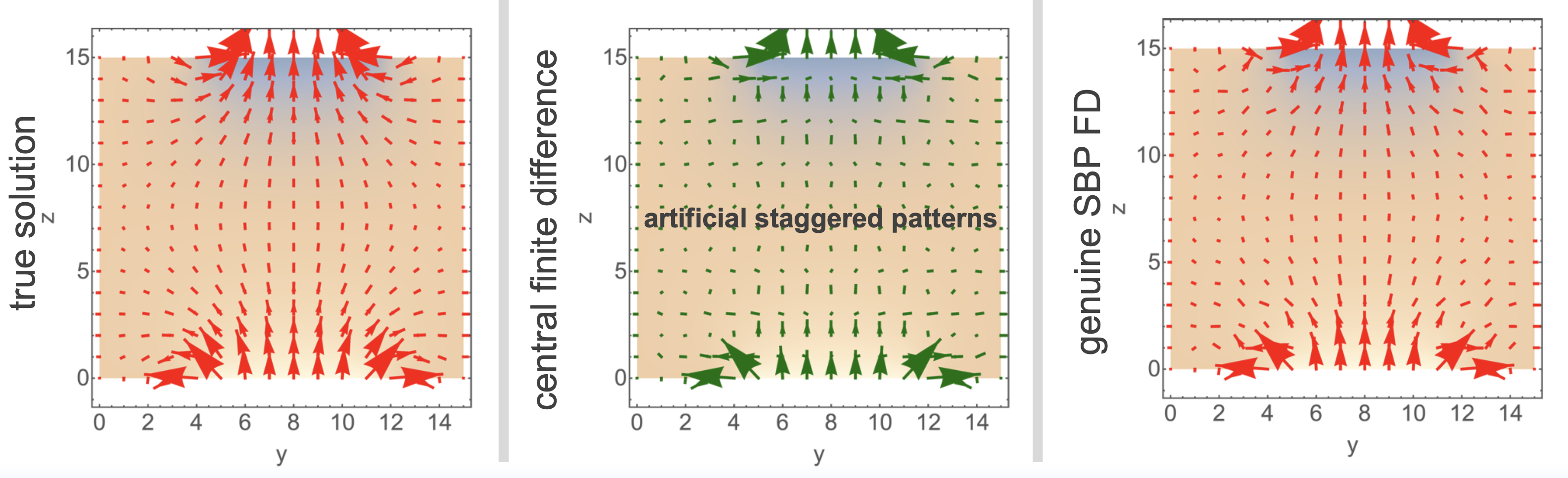}
\caption{(left) True field solution of the circular capacitor in red. (center) Central finite difference solution (green) of Gauss' law with the true solution as boundary condition, (right) SBP solution of Gauss' law in red.}\label{Fig:AbelianGaussCap2}
\end{figure}

Going over to a central finite difference discretization of Gauss' law appears to be a straight forward solution to this issue. However, as shown in the center panel of \cref{Fig:AbelianGaussCap2}, this choice of discretization, while producing symmetric field lines, introduces an artificial staggered pattern among them that does not yield an accurate reproduction of the true solution.

The correct way to treat non-trivial boundary conditions with finite differences requires instead a consistent treatment of integration and differentiation embodied in the concept of summation by parts (SBP). One aims to construct a finite difference operator (see e.g. \cite{2014JCoPh.26817S}) to fulfill this discrete counterpart to integration by parts.

To be concrete: if we discretize the inner product on a real function space $\int_0^L dx f(x)g(x) \approx {\cal T}_0^N[f_xg_x]$, we require the finite difference operator to fulfill 
\begin{align}
  {\cal T}_0^N[ (\Delta^{\rm SBP}f_x)g_x]   \overset{!}{=} -{\cal T}_0^N[ f_x (&\Delta^{\rm SBP}g_x)] + f_{N}g_{N}- f_0g_0.
\end{align}
The simplest form to implement an SBP finite difference for a first order derivative is given by the following example: using the matrix $H=a \; {\rm diag}[\frac{1}{2},1,\ldots,1,\frac{1}{2}]$ to denote the inner product between the two vectors of discretized functions $f=[f(0),f(a),\ldots,f((N_x-1) a)]$, the corresponding lowest order SBP operator for $N_x=4$ reads
\begin{align}
\Delta^{\rm SBP}=a^{-1}\left[ \begin{array} {cccc} -1 & 1& 0 &0 \\ -\frac{1}{2} & 0 & \frac{1}{2} &0 \\ 0 & -\frac{1}{2} & 0 & \frac{1}{2} \\ 0 &0 & -1 & 1 \end{array} \right].
\end{align}
And indeed, as can be seen in the right panel of \cref{Fig:AbelianGaussCap2} the combination of forward and backward FD on the boundary together with central FD in the interior leads to a much more accurate reproduction of the field lines for the Abelian capacitor example.

\begin{figure}
\centering
\includegraphics[scale=0.25]{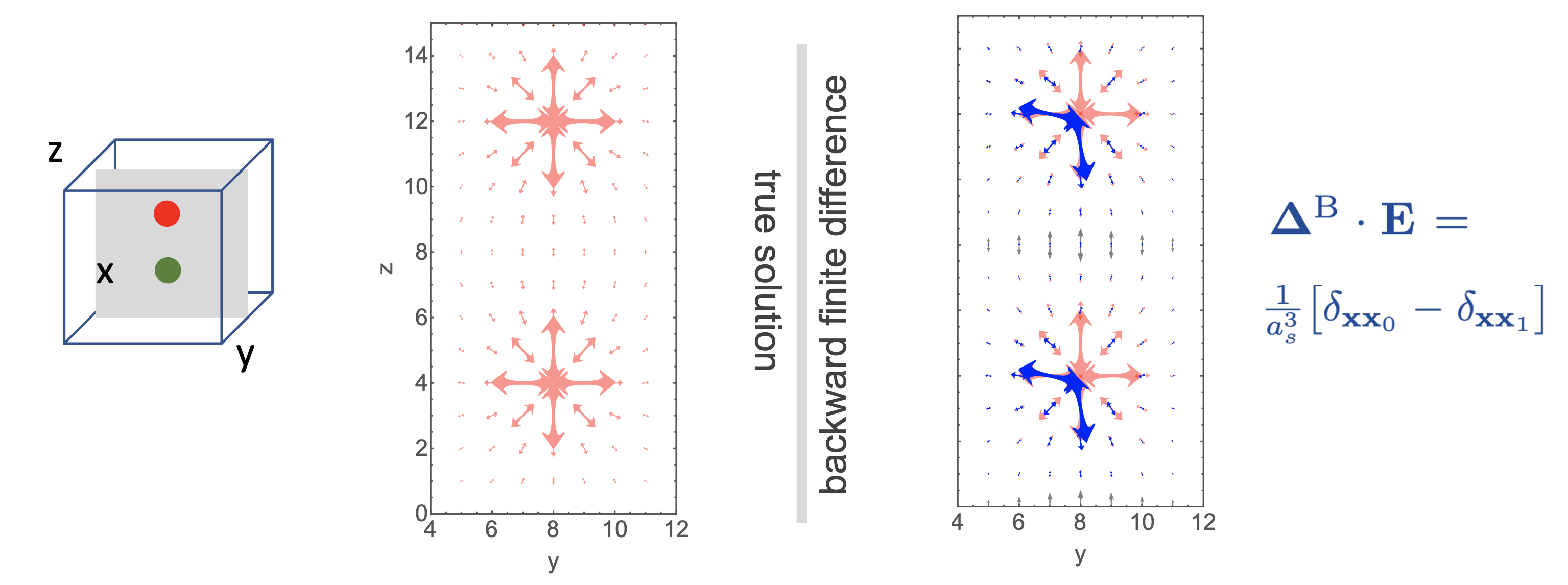}
\caption{(left) True solution of the eigenvectors of the stress tensor in the presence of a static charge anti-charge pair (right) solution of the corresponding backward FD Gauss' law is given as blue arrows.}\label{Fig:GaussLawChargesI}
\end{figure}

The second challenge is related to the accurate reproduction of field lines on the level of the stress tensor, the spatial components of $T^{\alpha\beta}=\frac{1}{4\pi}\big( g^{\alpha\mu}F_{\mu\lambda}F^{\lambda\beta}+\frac{1}{4}g^{\alpha\beta}F_{\mu\lambda}F^{\mu\lambda}\big)$. As concrete example I show a pair of a static charge and anti-charge placed at a certain distance in the left panel of \cref{Fig:GaussLawChargesI}. Diagonalizing the $3\times3$ spatial submatrix of $T^{\alpha\beta}$ yields a set of eigenvectors, two of which due to symmetry lie in the plane shown in \cref{Fig:GaussLawChargesI}, one is perpendicular to it. The in-plane eigenvector associated with positive eigenvalues is plotted and encodes the direction and strength of the field lines between the charges.

\begin{figure}
\centering
\includegraphics[scale=0.25]{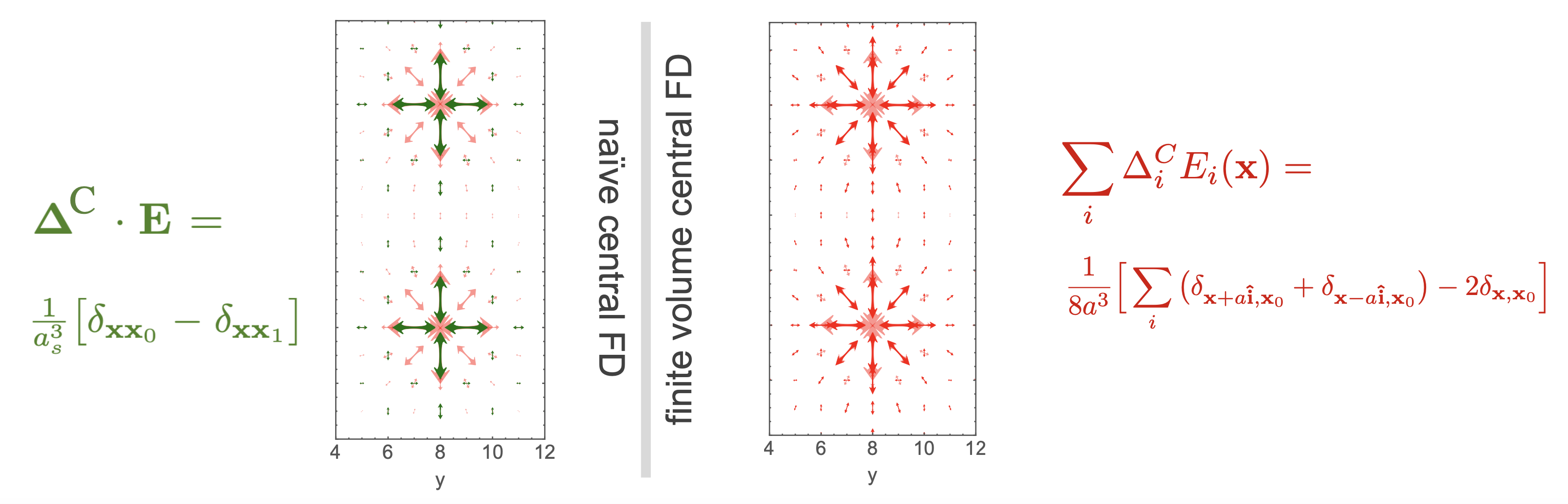}
\caption{(left) Solution of the central FD Gauss' law for a static charge anti-charge pair (right) solution of the corresponding finite volume discretized Gauss' law. In each panel the true solution is given in the background as light red arrows for comparison. }\label{Fig:GaussLawChargesII}
\end{figure}

The solution of the same scenario based on the backward FD is shown on the left and one immediately sees the asymmetry it introduces. Again we are in need of a symmetric discretization, however it is a known shortcoming of central finite differences to not respect the integral form of Gauss' law, i.e. in general
\begin{align}
Q=\int \, dV\, q = \int \, dV \,  {\bf \nabla^{\rm FD} E} \neq \int_{\partial V} \, d{\bf A} \cdot {\bf E}.
\end{align}
The artifacts that this introduces are clearly visible among the green arrows in the left panel of \cref{Fig:GaussLawChargesII}. We again encounter an artificially staggered pattern of the field lines.

A solution to this issue has been developed in the context of computational electrodynamics and it amounts to going over to a so called finite-volume discretization. The starting point of which is to consider the integral form of Gauss' law 
\begin{align}
\int_{x_{i-1/2}}^{x_{i+1/2}}dx\int_{y_{i-1/2}}^{y_{i+1/2}}dy&\int_{z_{i-1/2}}^{z_{i+1/2}}dz\big( \frac{d E_x}{dx}+\frac{d E_y}{dy}+\frac{d E_z}{dz}\big)= \int d^3x \delta^{(3)}({\bf x}-{\bf x}_0). \label{eq:FiniteVolDiscr}
\end{align} 
and to discretize its LHS using the midpoint rule. This expression is not in the form of a finite difference. It can be brought into that form if we combine multiple copies of it, each shifted by one lattice spacing in one of the spatial directions. The resulting expression 
\begin{align}
\sum_i \Delta_i^{\rm C} E_i({\bf x})&=\frac{1}{8a^3} \Big[ \sum_i\big( \delta_{{\bf x}+a\hat{\bf i},{\bf x}_0} +\delta_{{\bf x}-a\hat{\bf i},{\bf x}_0}\big)-2\delta_{{\bf x},{\bf x}_0}\Big] \label{eq:ConsGL}
\end{align}
indicates that the central finite difference needs to be amended with a particular form of distributed sources to fulfill Gauss' law. The resulting much improved reproduction of the field lines is shown as red arrows on the right of \cref{Fig:GaussLawChargesII}.

\section{Solving the two challenges in a non-Abelian setting}

In order to achieve a summation-by-parts discretization of lattice gauge theory, we require a central finite difference scheme for the interior of our system. Since the discretization there is correct up to ${\cal O}(a^2)$ we need to use a discretization for the links of the same order $\bar U_{\mu,x}={\rm exp}\big[ia_\mu A_{\mu,x+\frac{1}{2}a\hat\mu}\big]={\rm exp}\big[ia_\mu\frac{1}{2}\big(A_{\mu,x}+A_{\mu,x+a\hat\mu}\big)\big]+{\cal O}(a^2)$. As put forward in a recent preprint \cite{Rothkopf:2021jye} I propose to achieve this type of symmetric discretization by considering a $2\times2$ plaquette centered around the spacetime point $x$, which reads
\begin{align}
\nonumber P^{2\times2}_{\mu\nu,x}=& \bar U_{\mu,x-a\hat\mu-a\hat\nu}\bar U_{\mu,x-a\hat\nu}\bar U_{\nu,x+a\hat\mu-a\hat\nu}\bar U_{\nu,x+a\hat\mu}\bar U^\dagger_{\mu,x+a\hat\nu}\bar U^\dagger_{\mu,x-a\hat\mu+a\hat\nu}\bar U^\dagger_{\nu,x-a\hat\mu}\bar U^\dagger_{\nu,x-a\hat\mu-a\hat\nu}\\
\nonumber =&{\rm exp}\big[ 4ig a_\mu a_\nu \bar F_{\mu\nu,x}\big] + {\cal O}(a^3), \quad {\rm and \, where} \quad 
\bar F_{\mu\nu,x}= {\bf \Delta}^{\rm C}_\mu A_{\nu,x} - {\bf \Delta}^{\rm C}_\nu A_{\mu,x}+i[A_{\mu,x},A_{\nu,x}].
\end{align}
This type of plaquette can now be used to form a gauge invariant classical action for lattice gauge theory in the interior of the system
\begin{align}
 S^{2\times 2}=\sum_{x\notin \partial V} a_ta_s^3\Big[ \frac{2}{16 a_t^2a_s^2}\sum_i{\rm ReTr}\big[ 1-P^{2\times2}_{0i,x}\big]
 - \frac{1}{16 a_s^4}\sum_{ij} {\rm ReTr}\big[ 1-P^{2 \times2}_{ij,x}\big] \Big].
\end{align}
Since the action $S^{2\times 2}$ respects the gauge invariance of the lattice theory and reduces to the correct action in the continuum limit, it is a viable candidate to quantize the theory via the path integral framework. In order to achieve a genuine SBP discretization of the field strength tensor and in turn of Gauss' law, we need to amend the action $S^{2\times 2}$ on the boundaries with expressions that are based on $2\times 1$ and $1\times 2$ plaquettes, where as in the corners the $1 \times 1$ plaquettes are deployed.

Actions based on plaquettes with more than unit area have been considered before in the literature in the context of Symanzik's improvement program. I would like to note that in that context $2\times 2$ plaquettes were used predominantly in combination with the $1\times 1$ plaquette action and they were considered as oriented in the forward direction. In that fashion they did not realize the symmetry around the spacetime point $x$ of the present proposal and also do not lend themselves to constructing a genuine summation by parts discretization that captures non-trivial boundary conditions.

Having discussed a possible solution to the challenge of non-periodic boundary conditions, the remaining challenge amounts to implementing the finite volume discretization of Gauss' law in the interior of the volume. This, as discussed in the previous section, is necessitated by the presence of a central finite difference discretization in the SBP action. While in classical electrodynamics, formulated on the level of Maxwell's equations, the discretization of Gauss' law may be simply changed by hand, for the quantum theory we need to formulate an appropriate action. I.e. we need to introduce the spatial distribution of the sources in the fermion action.

As a first step let me restrict the discussion to heavy fermions where only a temporal covariant derivative acts on the spinors
\begin{align}
S_\psi=a_ta_s^3\sum_x i \psi_x^\dagger D_0 \psi_x.
\end{align}
In the classical equation of motion, the heavy fermions enter as charge density dynamically via their equal time commutator $J_0=\frac{g}{2}{\rm Tr}\left[\langle [\psi_x^\dagger,\psi_x]\rangle\right]$ as shown in Ref.\cite{Kasper:2014uaa}. Hence in order to realize the staggered pattern of \cref{eq:ConsGL}, shifts in spatial directions need to be realized in the covariant derivative. My proposal is to take inspiration from a finite difference operator  $\Delta^{\rm RN-SBP}_{t(j)}\phi(x)=(\phi(x+a\hat0-a\hat j)-\phi(x-a\hat0-a\hat j)-\phi(x-a\hat0+a\hat j)+\phi(x+a\hat0+a\hat j) )/4a$, recently developed in the context of open quantum systems \cite{Alund:2020ctu}. It displaces the fermions not only in temporal but also in spatial direction and when amended by appropriate factors of the link variables can be made into a gauge invariant and symmetric expression that can serve as operator in the heavy fermion action
\begin{align}
S_\psi&=a_ta_s^3\frac{i}{3} \sum_{x,i}  \psi_x^\dagger \bar D_{0(i)} \psi_x= a_ta_s^3\frac{i}{3} \sum_{x,i}  \psi_x^\dagger\frac{1}{8a}\Big[ 
\big( U_{0,x} U^\dagger_{j,x+a\hat 0-a\hat j}+ U^\dagger_{j,x-a\hat j}U_{0,x-a\hat j}\big)\psi_{x+a\hat 0-a\hat j}\\
\nonumber -&\big( U^\dagger_{j,x-a\hat j}U^\dagger_{0,x-a\hat 0-a\hat j}+ U^\dagger_{0,x-a\hat 0}U^\dagger_{jx-a\hat 0-a\hat j}\big) \psi_{x-a\hat 0-a\hat j}-\big(U^\dagger_{0,x-a\hat 0}U_{j,x-a\hat 0}+U_{j,x}U^\dagger_{0,x-a\hat0+a\hat j}\big) \psi_{x-a\hat0+a\hat j}\\
\nonumber +&(U_{j,x}U_{0,x+a\hat j}+U_{0,x}U_{j,x+a\hat 0}\big)\psi_{x+a\hat 0+a\hat j} \Big].
\end{align}
\section{Summary}
The question of quark-gluon-plasma production in small collision systems, as well as the precision exploration of the complex potential acting between a heavy quark-antiquark pair requires a consistent formalism for lattice gauge theory in the absence of (hypercubic-)translational invariance. On the level of the discretized classical action, the treatment of Dirichlet boundary conditions requires finite difference approximations that respect the summation by parts property, which I here propose to implement in the non-Abelian case via a novel centered $2\times2$ plaquette action in the interior and appropriate combinations of $1\times 2$, $2\times 1$ and $1\times 1$ plaquettes on the boundary. In order to preserve Gauss' law in the presence of a central finite difference discretization in the interior I take inspiration from computational electrodynamics to propose a finite volume discretiztion, which translates to spatially shifted sources in the finite difference language. A reparametrization invariant finite difference discretization is put forward that introduces the appropriate spatial shifts, when deployed in the action of heavy fermions.

\end{document}